\documentclass[11pt,a4wide]{article}

\usepackage{amsfonts}
\usepackage{amsbsy}
\usepackage{epsfig}
\usepackage{latexsym}
\usepackage{pst-tree}
\input amssym.def
\input amssym.tex

\advance \topmargin by -\headheight
\advance \topmargin by -\headsep
\evensidemargin \oddsidemargin
\advance\hoffset by -3mm 
\advance\voffset by 8mm 
\def\bbbc{{\mathchoice {\setbox0=\hbox{$\displaystyle\rm C$}\hbox{\hbox
to0pt{\kern0.4\wd0\vrule height0.9\ht0\hss}\box0}}
{\setbox0=\hbox{$\textstyle\rm C$}\hbox{\hbox
to0pt{\kern0.4\wd0\vrule height0.9\ht0\hss}\box0}}
{\setbox0=\hbox{$\scriptstyle\rm C$}\hbox{\hbox
to0pt{\kern0.4\wd0\vrule height0.9\ht0\hss}\box0}}
{\setbox0=\hbox{$\scriptscriptstyle\rm C$}\hbox{\hbox
to0pt{\kern0.4\wd0\vrule height0.9\ht0\hss}\box0}}}}

\makeatletter
\@addtoreset{equation}{section}
\makeatother

\begin{document}

\hfuzz=100pt \title{{\Large \bf{Pseudo-Riemannian Ricci-flat and Flat Warped Geometries and New Coordinates for the Minkowski Metric}}}
\author{\\M M Akbar\footnote{E-mail: akbar@utdallas.edu} \\
Department of Mathematics,\\ University of Texas at Dallas, \\ Richardson, TX 75080, USA\\
\&\\
School of Mathematics and Statistics,
\\ University of New South Wales, \\ Sydney, NSW 2052, Australia} \date{\today} \maketitle
\begin{abstract}
  \noindent It is well-known that the Einstein condition on warped
  geometries requires the fibres to be necessarily Einstein. However, exact warped solutions have
  often been obtained using one- and two-dimensional bases. In this paper, keeping the dimensions and signatures of the base and the fibre independently arbitrary, we obtain all Ricci-flat warped metrics with flat base in closed form and show that the number of free parameters is one less than the dimension of the base.
  Without any assumptions on the base and fibre geometry, we then show that a warped geometry is flat, i.e, has vanishing Riemann curvature, only if its base is flat and its fibre is maximally symmetric, i.e. of constant curvature.
     Applying this result systematically all possible warped forms of the Euclidean, Minkowski, and flat metrics of arbitrary signature can be obtained in closed form up to disjoint diffemorphisms of the base and  fiber metrics.
    In particular,  we obtained four new time-dependent forms of the Minkowski metric in four dimensions in addition to reproducing all of its known warped forms.
 \end{abstract}
\section{Introduction}
If $\cal{B}$ and $\cal{F}$ are two manifolds, with metrics $\overline{g}_{ab}$ and
$\widetilde{g}_{\mu\nu}$ respectively, and $f : {\cal{B}} \rightarrow {\mathbb{R}^{+}}$, their warped product ${\cal{B}}{\times_f} \,{\cal{F}}$ is the manifold, ${\cal{M}}$, with the
metric $G=\overline{g} \oplus f^2 \widetilde{g}$.  $\cal{B}$ and $\cal{F}$ are respectively called the base and fibre of the warped geometry. Although first systematically studied by O'Neill \cite{ONeill} in mid-sixties, examples of warped
geometries are ubiquitous, especially in the case of a one dimensional base. Simplest among them perhaps is the flat metric on $\mathbb{R}^d-\{0\}$ as $(d-1)$-dimensional sphere fibred over $\mathbb{R}^1$ with $f\equiv r>0$:
\begin{equation}
ds^2=dr^2+r^2(\mathbb{S}^{d-1}),
\end{equation}
well-known from the polar coordinate transformations of the standard Euclidean metric. A less well-known example perhaps is that of the Minkowski metric written as a Lorentzian cone over the hyperbolic space:
\begin{equation}
ds^2=-dt^2+t^2(\mathbb{H}^{d-1}).
\end{equation}
Standard cosmological models and other cohomogeneity-one Lorentzian, Riemannian or arbitrary signature semi-Riemannian metrics are warped products. They are the simplest warped geometries in that the base is a one dimensional line.

Warped geometries that are Einstein have naturally been looked into with particular interest.
In general, for the product geometry  $({\cal{M}},G)$ to be Einstein, i.e. $R=\Lambda G$, where $R$ is Ricci curvature tensor of $G$,  for some constant $\Lambda$, the
fibre $({\cal{F}}, {\widetilde{g}})$ has to be Einstein too, ${\widetilde{R}}=\lambda {\widetilde{g}}$ \cite{Besse}. With a one dimensional
base, it is easy to integrate for the warping function $f$ whose functional form depends on the signs of $\Lambda$ and $\lambda$. This can be done for both Lorentzian and Riemannian signatures.

However, since a two dimensional manifold can admit curvature it becomes impossible to obtain the warping function exactly for Einstein warped products \cite{Besse}.  This perhaps explains the lack of interest in finding exact solutions with base dimensions two and higher\footnote{Exact solutions for two-dimensional base however are still possible if one imposes restrictions on the metric of the base \cite{Besse, Petersen}.}. However, based on various topological and/or completeness assumptions, many powerful results have been obtained for Riemannian warped product spaces, including those that are Einstein (see, for example, \cite{Petersen} and references therein). For pseudo-Riemannian warped products see \cite{Beem, ONeill}.

In this paper we first note that a large class of exact solutions went unnoticed, namely when the base is flat and the warped geometry is Ricci-flat. No assumptions will be made for the dimensions and signature of the base and fibre manifold, so the solutions are pseudo-Riemannian thus encompass Riemannian and Lorentzian cases. We will see that it is possible to solve for all such geometries with arbitrary dimensions for the fibre as well as the base. 

 We then  study warped geometries that are flat (Riemann-flat, i.e., have vanishing Riemann curvature tensor) without any assumptions on the base and the fibre geometries or their signatures and find that warped geometries have vanishing Riemann tensor only if the base is flat and the fibre is a maximally symmetric (Einstein) space. We find the most general warped-product form of flat metrics of arbitrary signature in closed form up to separate coordinate transformations of the base and fibre (this constitutes a special case of the Ricci-flat solutions with decidedly flat-base above).\footnote{This paper does not deal with doubly-warped or multiply-warped metrics.} Our result allows us to systematically generate all warped forms of a flat metric of given dimension and signature explicitly. In the case of four dimensions, four surprising time-dependent forms of the Minkowski metric and one new form for the Euclidean metric are found, in addition to all their known warped forms.

\section{Einstein Warped Product}
In the following   $x^a, a=1...m$ and $y^{\mu}, \mu=1...n$ will denote coordinates of the base $({\cal{B}},{\overline{g}}_{ab})$ and the fibre $({\cal{F}}, {\widetilde{g}}_{\mu\nu})$, respectively, and an overline and tilde will represent base and fibre quantities, respectively, so the line element of the warped geometry  $({\cal{M}},G)$ is
\begin{eqnarray}
ds^2&=& G_{AB}\, dX^A dX^B\\
\,&=& {\overline{g}}_{ab}\, d{\overline{x}}^a d{\overline{x}}^b +f ^2({\bf{\overline{x}}}) {\widetilde{g}}_{\mu\nu}\,
dy^{\mu} dy^{\nu}, \label{warped}
\end{eqnarray}
in other words,
\begin{eqnarray}
G_{AB}&=&\left \{ \begin{array}{ll} \overline{g}_{ab}\ & {\rm if}\,\,1\le A,\,B \le m\ ,
\\ f^2 ({\bf{\overline{x}}}) {\widetilde{g}}_{\mu\nu}&{\rm if}\,\, m+1\le A,\,B \le d=m+n \ ,\\
0 &{\rm otherwise.} \end{array} \right. \label{warped1}
\end{eqnarray}
Its Ricci curvature tensor is easy to compute and given in the next section.
 As mention earlier, if one requires (\ref{warped}) to be Einstein (see, for example, \cite{Besse}) $R_{AB}=\Lambda G_{AB}$ then $({\cal{F}}, {\widetilde{g}_{\mu\nu}})$  must be Einstein as well, a result we will use in the rest of the paper without proof.
\subsection{Pseudo-Riemannian Ricci-flat warped metrics with flat base}
{\bf{Theorem 2.1}}: Let $\cal{B}$  be a $p+q=m$-dimensional flat space with $p$ negative and $q$ positive eigenvalues with the canonical metric $d\overline{s}^2=-dt_i^2+ dx_j^2$, $i=1,2,..,p; \,\,j=1,2,..,q$ on it and $\cal{F}$ is an Einstein space with metric $d\widetilde{s}^2=\widetilde{g}_{\mu\nu}dy^{\mu} dy^{\nu}$ of scalar curvature $\widetilde{R}=(n-1)\kappa$ where $\kappa =-1,0, \mathrm{or}\,\, 1$. Then all Ricci-flat warped product solutions ${\cal{B}}{\times_{f}} \,{\cal{F}}$ are given by the $(p+q-1)$-parameter
family
\begin{equation}
ds^2=-dt_i^2+ dx_j^2+(a_i t_i+ b_j x_j+c)^2 \widetilde{g}_{\mu\nu}dy^{\mu} dy^{\nu}, \label{mainmetric}
\end{equation}
where $a_i$ ($i=1,2,..,p)$, $b_j$ ($j=1,2,..,q)$ and $c$ are constants, satisfying
\begin{equation}
(n-1) (\Sigma a_i^2-\Sigma b_j^2+\kappa)=0.\label{cond}
\end{equation}
\\
{\bf{Proof}}: The components of the Ricci curvature tensor of (\ref{warped}) are:
\begin{eqnarray}
R_{a b} &=& \overline{R}_{ab}-\frac{n}{f}\, \overline{\nabla}_{a}\overline{\partial}_b f,\label{expression1}\\
R_{a\mu} &=& 0,\\
R_{\mu\nu} &=& \widetilde{R}_{\mu\nu} - {\widetilde{g}}_{\mu\nu}\left(f {\overline {\Delta}}f + (n-1) \overline{\nabla}^{a}f \overline{\nabla}_{a}f\right),\label{expression2}
\end{eqnarray}
The constant scalar curvature of the fiber and the flatness of the base imply
\begin{eqnarray}
\overline{\nabla}_a\overline{\partial}_b f &=& 0\,\, \forall\, a, b\label{expression111}\\
(n-1) \left(\overline{g}^{ab} \overline{\partial}_{a}f \overline{\partial}_{b}f-\kappa\right)&=& 0.\label{expression211}
\end{eqnarray}
Since the metric on the flat base in the chosen coordinates has vanishing Christoffel symbols, (\ref{expression111}) implies $f({\bf{\overline{x}}})$ is a linear function of its coordinates: $f({\bf{\overline{x}}})=a_i t_i+ b_j x_j +c$ which on substitution into (\ref{expression211}) gives (\ref{cond}). In general the parameter $c$ can be absorbed by shifting one of the base coordinates (unless all $a_i$ and $b_j$ are identically zero in which case (\ref{mainmetric}) is a trivial product), and (\ref{cond}) reduces the number of free parameters  to $(p+q-1)$ for $n > 1$.
$\Box$
\\
\\
{\it{Remark 2.1:}} When the base has exclusively positive or negative eigenvalues ($p=0$ or $q=0$) one necessarily needs $\kappa>0$ and $ \kappa < 0$ respectively for (\ref{cond}) to be satisfied. Otherwise $\kappa$ can have arbitrary signature.
\\
\\
{\it{Remark 2.2:}} Trivial product $a_i=b_j=0, c\ne 0$ is possible only if $\kappa=0$, in which case shift or any transformation of base coordinates cannot absorb $c$.
\\
\\
{\it{Remark 2.3:}} The signature of the fibre is arbitrary, thus the number of negative and positive eigenvalues of the warped product depend on the base and the fibre.
\section{All Flat Warped Metrics}
By direct computation one can show that components of the Riemann curvature tensor of (\ref{mainmetric}) vanish if the fibre is a maximally symmetric (i.e. of constant curvature) Einstein spce\footnote{Such metrics are often referred to as homogeneous and isotropic.}. In fact we are able to prove a more general result without any restrictions on the base.\\
\\
{\bf{Theorem 2.2}}: A warped metric is Riemann-flat only if the base is flat and the fibre is an Einstein space of constant curvature i.e. a maximally symmetric space.
\\
\\
{\bf{Proof}}: For (\ref{warped}), the non-zero components of the Christoffel symbols are $\Gamma^a_{bc}=\overline{\Gamma}^a_{bc}$,  $\Gamma^a_{\mu\nu}= -\overline{g}^{ab} f f_{,a}
\, \widetilde{g}_{\mu\nu}$, $\Gamma^\mu_{b \nu}=\frac{f_{,b}}{f}\, \widetilde{g}_{\mu\nu}$ and  $\Gamma^\sigma_{\mu\nu}= \widetilde{\Gamma}^\sigma_{\mu\nu}$
from which the non-zero components of the Riemann curvature tensor are found to be:
\begin{eqnarray}
R_{abcd}&=&\overline{R}_{abcd}, \label{riem1}\\
R_{\mu bc\nu}&=& \left( \frac{1}{2}\overline{\nabla}_{b} \overline{\partial}_c f^2- \partial_b f\, \partial_c f   \right)\widetilde{g}_{\mu\nu}\label{riem2}\\
R_{\mu\nu\rho\sigma}&=& f^2\, \widetilde{R}_{\mu\nu\rho\sigma} + f^2\, \overline{g}^{ab}f_{,_a} f_{,_b}(\widetilde{g}_{\mu\sigma}\widetilde{g}_{\nu\rho}-\widetilde{g}_{\mu\rho}\widetilde{g}_{\nu\rho}).\label{riem3}
\end{eqnarray}
From (\ref{riem1}), the vanishing of $R_{abcd}$ implies the vanishing of $\overline{R}_{abcd}$, requiring the base to be flat.
Choosing Cartesian coordinates on the base, i.e. the line element $d\overline{s}^2=-dt_i^2+dx_j^2$, the Christoffel symbols vanish. The vanishing of  $R_{\mu bc\nu}$ then means $\partial_b \partial_c f=0, \forall \,b, c$, implying $f({\bf{\overline{x}}})$ is a linear function of the base coordinates: $f({\bf{\overline{x}}})=a_i t_i+ b_j x_j +c$. The vanishing of $R_{\mu\nu\rho\sigma}$ in (\ref{riem3}) for $n=1$ is trivial and for $n > 1$ implies
\begin{equation}
f^2\, \widetilde{R}_{\mu\nu\rho\sigma} + f^2\, (-\Sigma a_i^2+\Sigma b_j^2 ) (\widetilde{g}_{\mu\sigma}\widetilde{g}_{\nu\rho}-\widetilde{g}_{\mu\rho}\widetilde{g}_{\nu\rho})=0
\end{equation}
or
\begin{equation}
\widetilde{R}_{\mu\nu\rho\sigma}= (\Sigma a_i^2-\Sigma b_j^2 ) (\widetilde{g}_{\mu\sigma}\widetilde{g}_{\nu\rho}-\widetilde{g}_{\mu\rho}\widetilde{g}_{\nu\rho})
\end{equation}
for $f\ne 0$, thus requiring the fibre to be an Einstein space of constant curvature. $\Box$

Keeping with our convention for $R=(n-1)\kappa$, here
$\widetilde{R}_{\mu\nu\rho\sigma}=\kappa (\widetilde{g}_{\mu\sigma}\widetilde{g}_{\nu\rho}-\widetilde{g}_{\mu\rho}\widetilde{g}_{\nu\rho})$, giving us  $(n-1) (a_i^2-b_j^2+\kappa)=0$ with $f({\bf{\overline{x}}})=a_i t_i+ b_j x_j +c$. Thus we can make the following statement.
\\
\\
{\bf{Theorem 2.3}}:  Up to disjoint coordinate transformations of the base and fiber metrics, all warped-product flat metrics (of arbitrary dimension and signature) are given by
\begin{equation}
ds^2=-dt_i^2+ dx_j^2+(a_i t_i+ b_j x_j+c)^2 \widetilde{g}_{\mu\nu}dy^{\mu} dy^{\nu}\label{flatall}
\end{equation}
where $a_i$ ($i=1,2,..,p)$, $b_j$ ($j=1,2,..,q)$ and $c$ are constants, satisfying
\begin{equation}
(n-1) (\Sigma a_i^2-\Sigma b_j^2+\kappa)=0\label{cond1}
\end{equation}
and $\widetilde{R}_{\mu\nu\rho\sigma}=\kappa (\widetilde{g}_{\mu\sigma}\widetilde{g}_{\nu\rho}-\widetilde{g}_{\mu\rho}\widetilde{g}_{\nu\rho})$.
\\
\\
{\it{Remark 3.1:}} To cast the warped metric (\ref{flatall}) in complete Cartesian form, as is in principle possible for any flat metric, one would require non-trivial coordinate transformations that will mix the base and fibre coordinates.
\section{Flat Spaces: Minkowski, Euclidean and Others}
We will now discuss how to systematically work out all warped form of a flat metric, of any given dimension and signature, using the above results and explicitly work out all warped forms for the four-dimensional Minkowski and the Euclidean metrics.
\subsection*{{\it{General Generation Technique}}}
 In any given dimension $d$, to obtain a flat metric of arbitrary signature, allocate $m=1, 2, d-1$ dimensions to the base and choose the fibre to be a maximally symmetric space in the rest of the available dimensions, i.e $n=d-1, d-2,...., 1$. Note that (\ref{cond1}) only connects the positive and negative eigenvalues of the base with the curvature of the fibre and does not put any restriction/preference for the signature of the fibre. Therefore all the negative eigenvalues of the warped metric need not necessarily reside in the base. Thus one can allow the base to be a maximally symmetric pseudo-Riemannian space, as we will see below. We first note the following:
 \\
 \\
{\it{Remark 4.1}} For co-dimension one bases, i.e for one dimensional fibre, (\ref{cond1}) is satisfied trivially meaning $a_i$ and $b_j$ are all arbitrary. Thus one obtains
\begin{equation}
ds^2=-dt_i^2+ dx_j^2+(a_i t_i+ b_j x_j +c)^2dz^2 \label{onedbase}
\end{equation}
as a $(d-2)$ parameter flat metric (since $c$ can be eliminated by a shift and one of $a_i$ or $b_j$  can be absorbed by rescaling $z$). One can change the signature of any coordinate in this case including the fibre, thus allowing the signature to be arbitrary.
\\
\\
{\it{Remark 4.2}} For a fibre with vanishing curvature, but dimension $n>1$, (\ref{cond1}) implies $a_i^2=b_j^2$. Therefore no solutions exist if the the base metric has only positive or negative eigenvalues. With the observation above, the Euclidean metric cannot be expressed as a warped product of lower dimensional Euclidean spaces except when one of them (the fibre) is a line.
\subsection{Four dimensions: New forms for Minkowski and Euclidean metrics}
We will use the dimensions of the base to catalog the solutions. It has a certain advantage: one can merely imagine attaching a higher dimensional fibre with same $\kappa$ to the same base and obtain a higher dimensional solution. However this would not produce other solutions possible in that (higher) dimension where the base takes more of the available dimensions.
\subsubsection{Three dimensional base}
This is the same as in {\it{Remark 4.1}}. In four dimensions with Lorentzian/Riemannian signature (\ref{onedbase}) gives us:
\begin{equation}
ds^2=\pm dt^2+dx^2+dy^2+(at+bx+cy)^2 dz^2 \label{form4}
\end{equation}
where $a, b, c$ are all arbitrary. One of them however can be absorbed by redefining $z$. By mere change of variables equation (\ref{form4}) be written as
\begin{equation}
ds^2=(ax+by+cx)^2 dt^2+dx^2+dy^2+ dz^2 \label{form4.1E}
\end{equation}
and
\begin{equation}
ds^2=- (ax+by+cx)^2 dt^2+dx^2+dy^2+ dz^2. \label{form4.1}
\end{equation}
The latter is a generalization of Rinder's coordinates for Minkowski space (which is obtained by setting $b=c=0$ in (\ref{form4.1})). Note that the arbitrariness in $a, b, c$ allows the metric to admit any number of negative eigenvalues.
\subsubsection{Two dimensional base}
 For $n\ge 2$, condition(\ref{cond1}) has to be satisfied non-trivially. This requires $a_i^2=b_j^2$ for a flat fibre. From {\it{Remark 4.2}} the possibility of having the Euclidean metric as a non-trivial warped product of two $\mathbb{E}^2$ spaces is eliminated leaving the direct product as the only possibility\footnote{This is one instance where $c$ in $f({\bf{\overline{x}}})$ has a non-trivial role.}. For the Minkowski metric, taking the two-dimensional fibre Riemannian, we obtain the following three forms:
\begin{equation}
ds^2=-dt^2 + dr^2 + (ar+ at)^2 (dx^2+ dy^2) \label{form1}
\end{equation}
\begin{equation}
ds^2= -dt^2+dr^2+\left(at+(1+a^2)^{\frac{1}{2}}\,r\right)^2 (d\theta^2 + \sin^2{\theta} d\phi^2) \label{form2}
\end{equation}
\begin{equation}
ds^2= -dt^2+dr^2+\left((1+a^2)^{\frac{1}{2}}\,t + a r\right)^2 (d\theta^2 + \sinh^2{\theta} d\phi^2) \label{form3}
\end{equation}
in all of which $a\in (-\infty, \infty)$. We now let the base be Riemannian and the fibre be a maximally symmetric two-dimensional Lorentzian metric. The absence of negative directions in the base means that all $a_i$'s should vanish and non-trivial solutions are possible only if the two parameters are related by $-b_1^2-b_2^2+ \kappa=0$, which requires $\kappa>0$. Two dimensional Lorentzian space with $\kappa>0$ is de Sitter. This gives
\begin{equation}
ds^2=dt^2+dr^2+\left(at+(1-a^2)^{\frac{1}{2}}\,r\right)^2\left(-d\theta^2 + \sinh^2{\theta} d\phi^2\right) \label{Minkowski-deSitter1}
\end{equation}
where $a\in (-1, 1)$. The fact that Minkowski metric can be written as ({\ref{form1}}) has been known from the days of Rosen and Einstein's work in connection with gravitational waves. However, none of the forms (\ref{form2}-\ref{Minkowski-deSitter1}) to the best knowledge of the author, was known before, except for the special case of $a=0$.
\subsubsection{One dimensional base}\label{analog}
In this case one either has $a^2+\kappa=0$ or $b^2-\kappa=0$ requiring $\kappa$ to be strictly negative or positive. It reproduces the following well-known forms of the Euclidean and Minkowski metrics as Euclidean or Lorentzian cones over spaces of constant curvature
\begin{equation}
ds^2=dr^2+r^2(\mathbb{S}^{3}).\label{analog1}
\end{equation}
and
\begin{equation}
ds^2=-dt^2+t^2(\mathbb{H}^{3})\label{analog2}
\end{equation}
and
\begin{equation}
ds^2=dt^2+t^2\left({\mathrm{d}}{\mathbb{S}}^{2,1}\right) \label{Minkowski-deSitter2}
\end{equation}
where ${\mathrm{d}}{\mathbb{S}}^{2,1}$ is the Lorentzian de Sitter space. They also show that higher dimensional generalizations for a one dimensional base is immediate in that one simply replaces the fibre by a dimensionally higher (constant-curvature) fibre.

\section{Conclusion}
Euclidean and Minkowski metrics in various coordinates are the most widely used metrics. Although they were written in various coordinates and some of their warped-product forms were known for many years, there was no systematic study looking for all of their warped-product forms. In this paper we investigated this in full generality and considered all flat metrics of arbitrary dimension and signature. We found that all flat warped geometry are maximally symmetric Einstein fibres over flat bases. We obtained the warping function explicitly by taking the base metric in Cartesian coordinates (which can be rewritten for any coordinate transformation one wishes to performs on the base). Then, up to disjoint coordinate transformations of the base and fibres, all flat warped geometries are given by the general form in {\bf{Theorem 2.3}}.

Applying the theorem systematically to four dimensions we obtained all possible warped forms of the Minkowski and Euclidean metrics explicitly. Among them are four novel forms of the Minkowski metric (\ref{form4.1}), (\ref{form2}), (\ref{form3}) and (\ref{Minkowski-deSitter1}) and one for the Euclidean metric, (\ref{form4.1E}). The first one is a generalization of the Rindler coordinates. A large literature exists on Rindler's coordinates, which describes an observer under constant acceleration moving in a certain direction in the Minkowski space. In particular, it has been found crucial in the study of (thermal) quantum gravitational-like effect the observer experiences. It would thus be interesting to explore the physical interpretation and ramifications of (\ref{form4.1}) with non-zero values of the parameters which will take us beyond the scope of this paper and is left for future work.

As for forms (\ref{form2}), (\ref{form3}) and (\ref{Minkowski-deSitter1}) of the Minkowski metric, they were known only for the special case of $a=0$. It is surprising that they were not found before despite the explicit maximal symmetry in their codimension two fibre. In particular, note that
(\ref{form2}) and (\ref{form3}) are {\it{time dependent}} as opposed to the $a=0$ case. They provide one-parameter time-dependent embeddings of  two-sphere and hyperbolic space in four dimensions. With the maximal symmetry in their fibres, they are likely to find many applications.  In particular it is possible that (\ref{form2}) and (\ref{form3}) will be useful in the context of gravitational waves as the known form (\ref{form1}) was.

Finally we would like to comment on {\bf{Theorem 2.1}} which constructs Ricci-flat solutions with Einstein fibre and flat base in arbitrary dimensions and signatures. For a two dimensional fibre, all Einstein metrics are of constant curvature, thus they reduce to flat solutions of {\bf{Theorem 2.3}}. For fibres of dimension three and higher it is possible to obtain bona fide Ricci-flat solutions with Einstein spaces of non-constant curvature. When the total dimension of the warped solution is four the latter is a Lorentzian or Euclidean cone over a three-dimensional Einstein metric. But that such a warped metric is well-known to be Ricci-flat. {\bf{Theorem 2.1}} therefore gives non-trivial warped Ricci-flat solutions, with parameter dependence, in five dimensions and higher for bases with dimensions $n\ge 2$. They should be of interest to those who work in various higher dimensional theories.
\section*{Acknowledgements}
The author  would like to thank Ivor Robinson, Istvan Ozsvath, Wolfgang Rindler, Gary Gibbons, Tony Dooley, Tibra Ali and Mahbub Majumdar for useful discussions. This work is dedicated to the memory of author's mother as sadaqaye jariyah for her.

\end{document}